\documentclass[aip,amsmath,amssymb,twocolumn,]{revtex4}                

\usepackage{textcase}
\usepackage{graphicx}
\usepackage{dcolumn}
\usepackage{bm}
\usepackage{amsmath,amsthm,amssymb,mathrsfs}

\begin{document}

\title{Self-Oscillation in Spin Torque Oscillator Stabilized by Field-like Torque}

\author{Tomohiro Taniguchi}
\author{Sumito Tsunegi}
\author{Hitoshi Kubota}
\author{Hiroshi Imamura}

\affiliation{ 
National Institute of Advanced Industrial Science and Technology (AIST), Spintronics Research Center, Tsukuba 305-8568, Japan}

\date{\today}%

\begin{abstract}
  The effect of the field-like torque on the self-oscillation of the magnetization 
  in spin torque oscillator 
  with a perpendicularly magnetized free layer was studied theoretically. 
  A stable self-oscillation at zero field is excited for negative $\beta$ 
  while the magnetization dynamics stops for $\beta=0$ or $\beta>0$, 
  where $\beta$ is the ratio between the spin torque and the field-like torque. 
  The reason why only the negative $\beta$ induces the self-oscillation was explained 
  from the view point of the energy balance between the spin torque and the damping. 
  The oscillation power and frequency for various $\beta$ were also studied 
  by numerical simulation. 
\end{abstract}

\maketitle



Self-oscillation of the magnetization 
in a nanomagnet excited by spin torque 
has attracted much attention 
for its potential application to 
spintronics devices such as 
microwave generators, 
magnetic sensors, 
and a recording and reading head of high-density hard disk drives 
\cite{kiselev03,rippard04,houssameddine07,bonetti09,kudo10,suto11,sinha11}. 
Recent development on the enhancement of the perpendicular magnetic anisotropy 
of the CoFeB ferromagnetic layer 
by adding a MgO capping layer \cite{yakata09,ikeda10,kubota12} enables us 
to realize a spin torque oscillator (STO) 
with a perpendicularly magnetized free layer 
and an in-plane magnetized pinned layer. 
It was previously shown that this type of STO 
produced a high emission power ($\sim$ 0.5 $\mu$W) 
with a narrow linewidth ($\sim$ 50 MHz) 
by this self-oscillation \cite{kubota13}. 
It should be noted that 
in Ref. \cite{kubota13}, 
a relatively large magnetic field (from 2 to 3 kOe) was applied to STO 
normal to the film plane, 
while the self-oscillation at zero field was interesting 
from a practical point of view. 
Although the emission power at zero field was investigated experimentally, 
the observed value was very low (typically, a few nW) \cite{zeng13}. 


Unfortunately, 
it was theoretically shown that 
the self-oscillation could not be excited 
in this type of STO 
in the absence of the applied field \cite{taniguchi13}. 
Above the critical current at which 
the spin torque destabilizes the perpendicularly magnetized initial state, 
the magnetization directly moves to 
the anti-parallel direction with respect to the pinned layer magnetization, 
and stops its dynamics. 
This conclusion was analytically shown 
by calculating the energy balance 
between the work done by spin torque 
and the energy dissipation due to the damping, 
and was confirmed by the numerical simulation 
of the Landau-Lifshitz-Gilbert (LLG) equation \cite{taniguchi13}. 



The purpose of this letter is to show that 
the previous conclusion is drastically modified 
by taking into account the field-like torque 
because the presence of such torque modifies 
the energy balance between the spin torque and the damping. 
The numerical simulation of the LLG equation shows that 
a stable self-oscillation can be realized for a negative $\beta$, 
where $\beta$ is the ratio between the spin torque and the field-like torque, 
as defined in Eq. (\ref{eq:LLG}) below. 
We also calculate the current dependences of the power and the oscillation frequency of STO 
at a finite temperature. 
The power significantly increases above the critical current when $\beta<0$, 
whereas it rapidly decreases when $\beta=0$ or $\beta>0$. 
Also, the oscillation frequency remains relatively high for $\beta<0$, 
whereas it drops to approximately zero for $\beta=0$ and $\beta>0$. 



\begin{figure}
\centerline{\includegraphics[width=0.7\columnwidth]{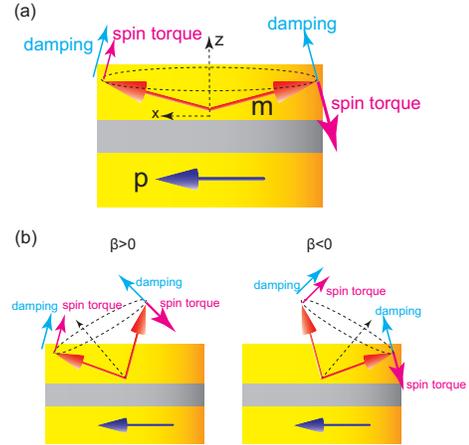}}\vspace{-3.0ex}
\caption{
           (a) Schematic view of the system. 
               The unit vectors pointing in the magnetization directions of 
               the free and the pinned layers are denoted as $\mathbf{m}$ and $\mathbf{p}$, respectively. 
               The directions of the spin torque and the damping are indicated by 
               the red and blue arrows, respectively. 
               The spin torque strength near the anti-parallel alignment of $\mathbf{m}$ and $\mathbf{p}$ is 
               larger than that near the parallel alignment 
               for $\lambda>0$, where $\lambda$ is defined in Eq. (\ref{eq:H_s}). 
           (b) Schematic view of the magnetization dynamics 
               in the presence of the field-like torque with $\beta>0$ or $\beta<0$. 
         \vspace{-3ex}}
\label{fig:fig1}
\end{figure}



Figure \ref{fig:fig1} (a) schematically shows the system under consideration. 
The unit vectors pointing in the magnetization directions of 
the free and pinned layers are denoted as 
$\mathbf{m}$ and $\mathbf{p}$, respectively. 
The $z$-axis is normal to the film plane 
whereas the $x$-axis is parallel to the pinned layer magnetization. 
The current flowing along the $z$-axis is denoted as $I$, 
where the positive current $I>0$ corresponds to 
the electrons flowing from the free layer to the pinned layer. 
As long as the current flows uniformly along the $z$-axis 
and the magnetization dynamics is well described by the macrospin model, 
the following result is applicable to both the nano-pillar and nano-contact. 
The pinned layer magnetization $\mathbf{p}$ is assumed to be fixed. 
The magnetization dynamics of the free layer are 
described by the LLG equation \cite{lifshitz80,gilbert04,slonczewski89,slonczewski96,slonczewski02}, 
\begin{equation}
\begin{split}
  \frac{d \mathbf{m}}{dt}
  =&
  -\gamma
  \mathbf{m}
  \times
  \mathbf{H}
  -
  \gamma 
  H_{\rm s}
  \mathbf{m}
  \times
  \left(
    \mathbf{p}
    \times
    \mathbf{m}
  \right)
\\
  &-
  \gamma 
  \beta 
  H_{\rm s}
  \mathbf{m}
  \times
  \mathbf{p}
  +
  \alpha 
  \mathbf{m}
  \times
  \frac{d \mathbf{m}}{dt},
  \label{eq:LLG}
\end{split}
\end{equation}
where $\gamma$ and $\alpha$ are 
the gyromagnetic ratio and the Gilbert damping constant, respectively. 
Throughout this letter, 
we consider the case of the zero applied field. 
The magnetic field $\mathbf{H}=(H_{\rm K}-4\pi M) m_{z} \mathbf{e}_{z}$ 
consists of the crystalline anisotropy field $H_{\rm K}$ 
and the demagnetization field $4\pi M$. 
Because we are interested in the perpendicularly magnetized free layer, 
$H_{\rm K}$ should be larger than $4\pi M$. 
The magnetic field can be defined as 
the derivative of the energy density 
$E=-M(H_{\rm K}-4\pi M)m_{z}^{2}/2$ with respect to the magnetization $M \mathbf{m}$, 
i.e., $\mathbf{H}=-\partial E/\partial (M \mathbf{m})$. 
The effect of the Oersted field generated by the current, 
which causes a non-uniform magnetization dynamics 
and whose order is less than or equal to 10 Oe for the experimental parameters \cite{kubota13}, 
is neglected 
because the experimental results on the oscillation properties of 
this type of STO was quantitatively reproduced by the macrospin model \cite{kubota13}. 
The second and third terms 
on the right-hand side of Eq. (\ref{eq:LLG}) represent 
the spin torque and the field-like torque, respectively. 
The spin torque strength, 
\begin{equation}
  H_{\rm s}
  =
  \frac{\hbar \eta I}{2e (1 + \lambda \mathbf{m}\cdot\mathbf{p}) MV}, 
  \label{eq:H_s}
\end{equation}
consists of the volume of the free layer $V$ 
and the spin torque parameters, $\eta$ and $\lambda$. 
The parameter $\eta$ corresponds to the spin polarization of the injected current, 
i.e., the spin polarization of the pinned layer, 
and $\lambda$ determines the dependence of the spin torque strength 
on the relative angle between the magnetizations, $\mathbf{m}$ and $\mathbf{p}$. 
The theoretical relation between $\lambda$ and the material parameters depend on the model 
\cite{slonczewski89,slonczewski96,slonczewski02,slonczewski05}. 
For example, Ref. \cite{slonczewski05} calculated 
the spin torque from the transfer matrix of a magnetic tunnel junction (MTJ), 
and and showed that $\lambda=\eta \eta^{\prime}$, 
where $\eta^{\prime}$ is the spin polarization of the free layer. 
The sign of $\lambda$ is positive (negative) 
when the MTJ shows the positive (negative) tunnel magnetoresistance. 
For the positive (negative) $\lambda$, 
the spin torque magnitude near the anti-parallel alignment of $\mathbf{m}$ and $\mathbf{p}$ is 
larger (smaller) than that near the parallel alignment \cite{slonczewski96}. 
The self-oscillation can be realized 
when the energy supplied by the spin torque balances with 
the energy dissipation due to the damping. 
The energy supplied by the spin torque can be both positive and negative, 
depending on the current direction and the sign of $\lambda$. 
The net energy supplied during a precession should be positive 
to excite a stable self-oscillation. 
In the present geometry with the positive current $I>0$, 
the spin torque dissipates energy when $m_{x}>0$ 
because it is parallel to the damping, 
whereas it supplies energy when $m_{x}<0$ 
because it is anti-parallel to the damping, 
as shown in Fig. \ref{fig:fig1} (a). 
Then, the sign of $\lambda$ should be positive 
to make the net energy supplied by the spin torque positive. 
On the other hand, 
the net energy supplied by the spin torque becomes positive by the negative current when $\lambda$ is negative. 
Then, only the positive (negative) current can excite the self-oscillation of the magnetization 
for the positive (negative) $\lambda$ \cite{rippard04,taniguchi13}. 


The dimensionless parameter $\beta$ in Eq. (\ref{eq:LLG}) 
is the ratio between the spin torque and the field-like torque. 
The origin of the field-like torque is the same as 
that of the spin torque, 
i.e., the transfer of the transverse spin angular momentum 
from the conduction electron to the free layer magnetization. 
While the magnitude of the field-like torque is negligible 
in a giant magnetoresistive (GMR) system \cite{zhang02,zwierzycki05}, 
in an MTJ it reaches a few tens of percents of 
the spin torque magnitude \cite{theodonis06,xiao08}. 
The value of $\beta$ has been experimentally measured 
by using the spin torque diode effect \cite{tulapurkar05,kubota08,sankey08}, 
although the spin torque diode effect can measure $\beta$ 
below the critical current only 
while the self-oscillation state is realized by the current above the critical current. 
Another method to estimate $\beta$ is 
the measurement of the magnetization switching probability in the thermally activated region \cite{li08,rippard11}, 
in which the effect of the field-like torque is regarded as an enhancement or a reduction 
of the switching barrier height, 
and the magnitude of $\beta$ is estimated from the bias dependence of the switching probability. 
Both the theoretical calculation and the experimental measurement have shown that 
the magnitude and sign of the field-like torque depend on 
the material parameters such as the exchange splitting of the conduction electrons of the free and pinned layers, 
the sample thickness, and the bias voltage 
\cite{zhang02,zwierzycki05,theodonis06,xiao08,tulapurkar05,kubota08,sankey08}. 
However, for simplicity, $\beta$ in this letter is assumed to be constant 
with respect to the bias voltage (current). 
Instead, we study the magnetization dynamics for various values of $\beta$. 
The value of $|\beta|$ estimated from the spin torque diode measurement of 
the in-plane magnetized CoFeB/MgO/CoFeB MTJ \cite{kubota08,sankey08} is on the order of $0.01-0.1$. 
Similar values of $\beta$ were found from 
the measurement of the switching probability \cite{li08,rippard11}. 



\begin{figure}
\centerline{\includegraphics[width=0.65\columnwidth]{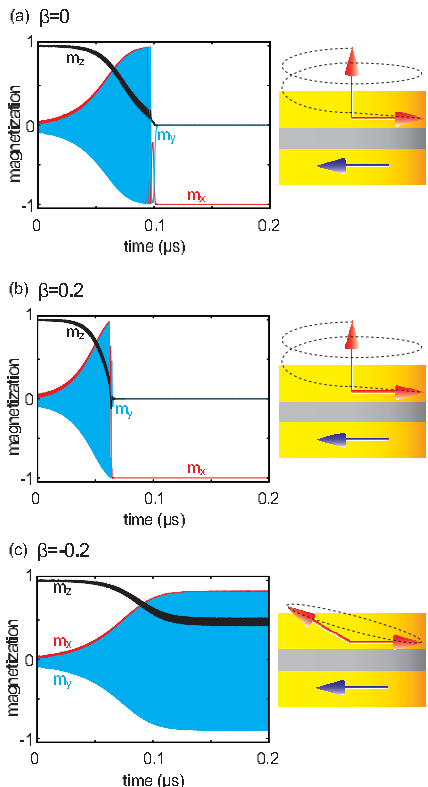}}\vspace{-3.0ex}
\caption{
        The time evolutions of the components of $\mathbf{m}$ 
        for (a) $\beta=0$, (b) $0.2$, and (c) $-0.2$, 
        where the red, blue, and black lines correspond to 
        $m_{x}$, $m_{y}$, and $m_{z}$, respectively. 
        The current magnitude is $1.5$ mA. 
        The magnetization stops at $\mathbf{m}=-\mathbf{p}$ for $\beta=0$ and $0.2$, 
        while it steadily precesses for $\beta=-0.2$. 
        The schematic views of the magnetization dynamics are shown on the right side. 
         \vspace{-3ex}}
\label{fig:fig2}
\end{figure}



Figures \ref{fig:fig2} (a), (b), and (c) show 
the time evolutions of the components of $\mathbf{m}$ 
obtained by numerically solving Eq. (\ref{eq:LLG}), 
in which the values of $\beta$ are 
(a) $\beta=0$, (b) $0.2$, and (c) $-0.2$, respectively. 
The schematic views of the magnetization dynamics are also shown 
on the right side. 
The values of the parameters are 
$M=1448$ emu/c.c., 
$H_{\rm K}=20.0$ kOe, 
$V=\pi \times 60 \times 60 \times 2$ nm${}^{3}$, 
$\eta=0.54$, 
$\lambda=\eta^{2}$, 
$I=1.5$ mA, 
$\gamma=1.732 \times 10^{7}$ rad/(Oe$\cdot$s), 
and $\alpha=0.005$, respectively \cite{yakata09,kubota12,kubota13,konoto13}. 
By using these parameters, 
the critical current to destabilize the initial state for $\beta=0$, 
$I_{\rm c}=[4 \alpha eMV/(\hbar \eta \lambda)](H_{\rm K}-4\pi M)$ \cite{taniguchi13}, 
is estimated to be $1.2$ mA, 
which corresponds to $11 \times 10^{6}$ A/cm${}^{2}$ in terms of the current density. 
As studied in Ref. \cite{taniguchi13}, 
in the absence of the field-like torque ($\beta=0$), 
the magnetization moves to the anti-parallel direction with respect to $\mathbf{p}=\mathbf{e}_{x}$, 
and stops its dynamics. 
Similarly, in the case of the positive $\beta$, 
the magnetization stops at $\mathbf{m}=-\mathbf{p}$. 
The convergence time of the magnetization for $\beta>0$ is 
shorter than that for $\beta=0$. 
Contrary to $\beta \ge 0$, 
stable self-oscillation is realized for negative $\beta$, 
as shown in Fig. \ref{fig:fig2} (c). 
The cone angle at the self-oscillation state increases 
as the current increases. 

\begin{figure}
\centerline{\includegraphics[width=1.0\columnwidth]{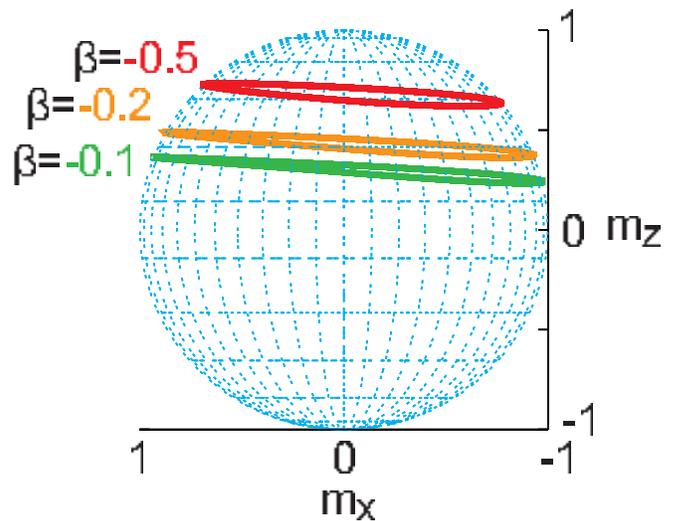}}\vspace{-3.0ex}
\caption{
         Trajectories of the steady state precession of the magnetization in the free layer 
         for various negative $\beta=-0.1,-0.2,-0.5$. 
         The current magnitude is $I=2.0$ mA. 
         \vspace{-3ex}}
\label{fig:fig3}
\end{figure}

The results shown in Fig. \ref{fig:fig2} indicate that 
the field-like torque plays a key role toward the realization of 
self-oscillation in this STO. 
As mentioned after Eq. (\ref{eq:H_s}), 
the self-oscillation can be realized 
when the energy supplied by the spin torque balances 
with the energy dissipation due to the damping. 
For $\beta=0$, 
in the absence of the applied field, 
spin torque above $I_{\rm c}$ always overcomes the damping 
during $m_{z}=1$ to $m_{z}=0$, 
whereby self-oscillation cannot be realized \cite{taniguchi13}. 
The magnetization moves to the film-plane 
and stops its dynamics at $\mathbf{m}=-\mathbf{p}$. 
In the case of $\beta \neq 0$, 
the field-like torque acts approximately as 
an applied field pointing in the positive (negative) $x$-direction 
for $\beta>0(<0)$, 
and modifies the precession trajectory, as shown in Fig. \ref{fig:fig1} (b), 
due to which the amount of energy supplied by 
the spin torque differs from that for $\beta=0$ \cite{taniguchi13a}. 
For the positive $\beta$, 
the amount of energy supplied by the spin torque increases, 
and the magnetization rapidly moves to $\mathbf{m}=-\mathbf{p}$, 
as shown in Fig. \ref{fig:fig2} (b). 
On the other hand, 
for the negative $\beta$, 
the energy supplied decreases. 
Then, the spin torque balances with the damping 
above the film-plane, 
and therefore, the stable self-oscillation of the magnetization can be realized. 
In Fig. \ref{fig:fig3}, 
the oscillation trajectories of the magnetization in the steady state 
for various negative $\beta=-0.1,-0.2,-0.5$ are shown, 
where the cone angle of the magnetization from the $z$-axis increases 
as the absolute value of $\beta$ decreases. 
This is because the amount of the energy supplied by the spin torque increases 
as $|\beta|$ decreases. 
It can also be seen from Fig. \ref{fig:fig3} that 
the oscillation trajectories are not centered around the $z$ axis. 



\begin{figure}
\centerline{\includegraphics[width=0.7\columnwidth]{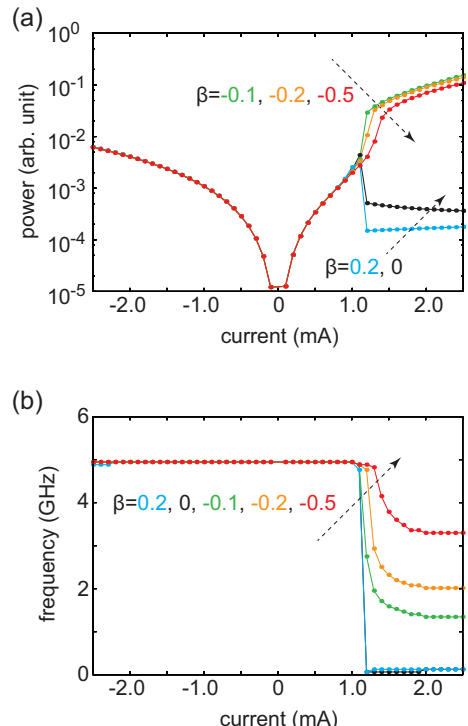}}\vspace{-3.0ex}
\caption{
         Dependences of (a) the powers and (b) the oscillation frequencies of STO 
         on the applied current 
         obtained from finite temperature simulations, 
         where the value of $\beta$ varies from $-0.5$ to $0.2$. 
         The current 1.0 mA corresponds to $8.8 \times 10^{6}$ A/cm${}^{2}$. 
         \vspace{-3ex}}
\label{fig:fig4}
\end{figure}



We also calculated 
the power and oscillation frequency of STO, 
both of which depend on $m_{x}(t)$ through 
the magnetoresistance effect $\propto \Delta R \mathbf{m}\cdot\mathbf{p}$ \cite{kubota13}, 
where $\Delta R=R_{\rm AP}-R_{\rm P}$ is the difference in the resistances 
between the parallel (P) and anti-parallel (AP) alignments of $\mathbf{m}$ and $\mathbf{p}$. 
The power is defined as 
$\int_{0}^{\infty} |m_{x}(f)| I^{2} df$ \cite{kubota13}, 
where $m_{x}(f)$ is the Fourier transformation of $m_{x}(t)$. 
The random torque, $-\gamma \mathbf{m} \times \mathbf{h}$, 
is added to the right-hand side of Eq. (\ref{eq:LLG}). 
This random torque excite a small amplitude oscillation of the magnetization, 
known as the mag-noise. 
Then, even if $\beta \ge 0$, 
the spectrum has a peak at the ferromagnetic resonance frequency, 
an the power can be evaluated, 
although a sustainable oscillation cannot be excited at zero temperature. 
The components of the random field $h_{k}$ ($k=x,y,z$) satisfy 
the fluctuation-dissipation theorem \cite{brown63}, 
$\langle h_{k}(t) h_{\ell}(t^{\prime}) \rangle = [2\alpha k_{\rm B}T/(\gamma MV)]\delta_{k \ell} \delta(t-t^{\prime})$, 
where $T$ is the temperature. 
The oscillation frequency is defined as the peak frequency of $|m_{x}(f)|$. 
The values of the parameters are those used in Fig. \ref{fig:fig2} 
with $T=300$ K. 
Because of the presence of the random torque, 
the trajectories of the magnetization dynamics include randomness. 
Therefore, we repeated the numerical simulation $10^{3}$ times with different random numbers, 
and averaged these spectra to calculate the power and oscillation frequency.


Figures \ref{fig:fig4} (a) and (b) show 
the powers and oscillation frequencies for various $\beta$, respectively. 
Only the positive current can induce the magnetization dynamics in this system \cite{taniguchi13}, 
and the power observed in the negative current region corresponds to the mag-noise power. 
We focus on the positive current region here. 
The most remarkable point is that 
the power significantly increases above the critical current ($\simeq 1.2$ mA) 
for $\beta<0$, 
whereas it decreases for $\beta=0$ and $\beta>0$. 
This is because, as shown in Fig. \ref{fig:fig2}, 
the stable self-oscillation of the magnetization can be realized 
in the case of $\beta<0$ 
whereas the magnetization dynamics stop at $\mathbf{m}=-\mathbf{p}$ 
in the cases of $\beta=0$ and $\beta>0$. 
As shown in Fig. \ref{fig:fig3}, 
the cone angle of the magnetization from the $z$-axis increases 
as the absolute value of $\beta$ decreases 
for negative $\beta$. 
Since the large amplitude of $m_{x}$ due to the large cone angle results the large power, 
the power for $\beta=-0.1$ is greater than that for $\beta=-0.5$. 
On the other hand, 
the oscillation frequency ($\simeq \gamma(H_{\rm K}-4\pi M)m_{z}/(2\pi)$)decreases 
as the cone angle of the magnetization, $\cos^{-1}m_{z}$, increases. 
Then, the oscillation frequency for $\beta=-0.1$ is lower than that for $\beta=-0.5$. 
In the cases of $\beta=0$ and $\beta>0$, 
the powers are the mag-noise powers 
originated from the random oscillation of the magnetization around 
the $z$- or $x$-axis, 
depending on whether the current magnitude is below or above the critical current. 
Because the oscillation amplitude of $m_{x}$ for the precession around the $z$-axis is 
larger than that for the precession around the $x$-axis, 
the power drops at the critical current. 
The oscillation frequency above the critical current is almost zero 
because the anisotropy field $(H_{\rm K}-4\pi M)m_{z}$ is zero 
in the $xy$-plane. 




In conclusion, 
the magnetization dynamics of STO 
with a perpendicularly magnetized free layer 
and an in-plane magnetized pinned layer was 
studied by numerically solving the LLG equation. 
We found that 
the field-like torque with negative $\beta$ enabled the realization of 
a stable large amplitude self-oscillation of the magnetization at zero field 
in this type of STO, 
where $\beta$ is the ratio between the spin torque and the field-like torque. 
On the other hand, the large amplitude oscillation at zero field 
cannot be excited for $\beta=0$ and $\beta>0$. 
From the thermally activated state below the critical current 
to the self-oscillation state slightly above the critical current, 
the power increases $10^{3}$ times for $\beta<0$, 
where the power is defined as 
the product of 
the integral of the Fourier spectrum of the oscillation amplitude and 
the square of the current. 
The oscillation frequency defined as the peak frequency of the Fourier spectrum 
remains on the order of a few GHz for $\beta<0$ 
as the current increases. 


The authors would like to acknowledge 
H. Maehara, A. Emura, T. Yorozu, S. Tamaru, H. Arai, M. Konoto, K. Yakushiji, T. Nozaki, A. Fukushima, K. Ando, and S. Yuasa. 
This work was supported by JSPS KAKENHI Number 23226001. 




\end{document}